\documentclass{emulateapj}

\def\kms{km~s$^{-1}$}

\def\Msun{\,M_\odot}

\shorttitle{Stellar Disruption and the Quasar Radio Dichotomy}
\shortauthors{Gopal-Krishna et al.}
\begin{document}

\title{Stellar Disruption by Supermassive Black Holes and the \\
Quasar Radio Loudness Dichotomy}

\author{Gopal-Krishna\altaffilmark{1},
    A.\ Mangalam\altaffilmark{2} and Paul J.\ Wiita\altaffilmark{3}}
\altaffiltext{1}{National Centre for Radio Astrophysics, TIFR, Post Bag 3, Pune Univ.\ Campus, 411007, India; krishna@ncra.tifr.res.in
}
\altaffiltext{2}{Indian Institute of Astrophysics, Sarjapur Road, Bangalore
560034, India; mangalam@iiap.res.in
}
\altaffiltext{3}{Department of Physics and Astronomy, Georgia State
University, Atlanta, GA 30302-4106; wiita@chara.gsu.edu
}


\begin{abstract}
The origin of the dichotomy of radio loudness among quasars can  be 
explained using recent findings that the mass of the central supermassive 
black hole (SMBH) in extended radio-loud quasars is systematically a few 
times that of their counterparts in radio-quiet quasars.  This sensitive
dependence of radio jet ejection upon SMBH mass probably arises from
the blockage of jets by the presence of substantial quantities of gas 
tidally stripped from stars by the central BH.  This disruptive gas, 
however, will only be available around BHs with masses less than 
$M_c ~\gtrsim ~10^8\Msun$, for which the tidal disruption radius lies 
outside the SMBH's event horizon.  Consequently, we find that AGN with
$M_{BH} > M_c$ can successfully launch jets with a wide range of powers,
thus producing radio-loud quasars.  The great majority of jets launched
by less massive BHs, however, will be truncated in the vicinity of the 
SMBH due to mass loading from this stellar debris. This scenario also can 
naturally explain the remarkable dearth of extended radio structures in 
quasars showing broad absorption line spectra.
\end{abstract}


\section{Introduction}

The bi-modality in the distribution of the radio-to-optical flux ratio,
$R$, was first revealed 
using optically selected samples
(e.g., Kellermann et al. 1989). 
A similar picture emerged from the plot of radio flux versus [O III] 
line intensity for QSOs (the latter marks
the accretion 
power of the central engine), which showed that radio loud quasars (RLQs) 
are typically $\sim 10^4$ times more radio luminous than radio quiet 
quasars (RQQs) at a given $L_{[O III]}$ (e.g., Miller, Rawlings \& 
Saunders 1990;  Xu et al.\ 1999).
However, a somewhat ambiguous view of this dichotomy emerged from 
more recent investigations based on deep, large sky area radio surveys 
(FIRST, Becker et al.\  1995; NVSS, Condon et al.\ 1998) and extensive 
optical surveys (SDSS, York et al.\  2000; 2dF, Croom et al.\ 
2001). Thus, while a radio loud fraction of $5-10\%$ was estimated in 
some studies (e.g., Ivezi{\'c} et al.\ 2002, 2004), other papers reported 
 no radio-loudness bimodality (White et al.\ 2000; 
Cirasuolo et al.\ 2003).
Laor (2003) has argued that these negative results are
probably an artifact of the lower sensitivity of the FIRST radio survey to
extended radio emission, which, unlike the core emission, is unbeamed, 
and hence a more reliable indicator of the jet power. In fact, a double 
peaked distribution of $R$, is seen in an ``image stacking" analysis 
of the FIRST data for QSOs (White et al.\ 2007). 

It is unlikely that the host galaxy morphology or the large-scale
environment play a crucial role in the radio loudness dichotomy since
in both these respects, optically bright RQQs are very similar to RLQs
(Metcalf \& Magliocchetti 2006, hereafter MM, and references therein). 
Therefore, it has been suggested that the main difference between
 RLQs and RQQs lies in the availability of 
SMBH spin energy to power the jets in RLQs 
(e.g., 
Wilson \& Colbert 1995; Hughes \& Blandford 2003). 
In the scheme of Sikora, Stawarz \& Lasota (2007), radio powerful jets,
driven by the SMBH spin,
are realized intermittently whenever a strong collimation is externally 
imposed on them by the MHD wind from the accretion disk (see, also, Nipoti 
et al., 2005).  
                                                                                
A key observational clue  for the radio  dichotomy may stem from
the inferred mass-related duality of radio loudness, in that powerful 
radio sources are only associated with galactic nuclei containing central
black holes of masses $M_{BH} > 10^8 M_{\sun}$ (Laor 2003; Dunlop 
et al.\ 2003; Chiaberge, Capetti \& Macchetto 2005; Sikora et al.\ 2007).  
Note that this clue does not exclude the spin of the black hole or
accretion disk from being the basic mechanism underlying 
jet formation. However, for radio detection the jet must be able to emerge 
from the nuclear region out to parsec scale. An ``aborted" jet scenario 
recently invoked by Ghisellini, Haardt \& Matt (2004) envisions
that radio-quiet
AGN too eject relativistic particle jets but they are terminated due
to the impact of infalling shells of ``heavy" material that are  
ejected by the AGN intermittently. 
A likely case of an intermittently aborted jet comes from the X-ray 
monitoring of the RQQ PG 1407+265 (Gallo 2006) which is believed to 
possess a highly Doppler boosted compact radio jet (Blundell et al.\ 2003). 

While nuclear jet abortion may offer an attractive route to explaining
radio quietness, can it be reconciled with the {\it dichotomy} of radio
loudness of quasars, which itself appears to be sensitively linked to
$M_{BH}$? 
Specifically, it needs
to be understood 
how the jet abortion scenario ties in with the observed 
systematic mass difference between the central SMBHs of RLQs and RQQs 
and the finding that all RLQs have $M_{BH} > 10^8 M_{\sun}$ (Laor 2003; 
see above). Several recent studies using different techniques and based 
on large-sky radio and optical surveys have revealed a consistent 
mass excess of 
the central black-holes powering RLQs vis-a-vis those in RQQs
(Dunlop et al.\ 2003; MM; Jarvis \& McLure 2006; Hyv{\"o}nen et al.\ 2007).
The determinations of $M_{BH}$ in these studies are based both on 
optical spectroscopy of the nuclei and accurate photometry of 
host galaxies using the Hubble Space Telescope. Moreover, care was
taken  that the samples of RQQs and RLQs 
are matched in 
redshift and optical luminosity. Thus, even though it is only by a modest 
factor (1.4 to 4), the presence of a mass excess of the BHs in RLQs over 
those in RQQs has been found in independent studies.  Clearly, if 
this is the main factor behind the radio dichotomy, the underlying 
physical process must be very sensitive to $M_{BH}$.  A possible link 
of the SMBH critical mass, $(M_c)$, to the radio dichotomy, via tidal
disruptions of stars, was noted in Laor (2000) but discounted on the 
ground that any influence of the tidal debris would be too short lived 
as the debris would be quickly sucked into the SMBH on a dynamical time 
scale (sub-day). Below we argue that even the modest difference 
found between the BH masses in RLQs and RQQs could play a critical role 
in aborting the nascent jets in RQQs.

\section{Disrupted Stars Can Choke Nascent Jets}

In our picture, twin-jets of relativistic plasma are driven
by the SMBH spin (e.g.\ Blandford \& Znajek 1977) or by the accretion 
disk (e.g., Blandford \& Payne 1982), and/or by variants involving a 
magnetic coupling between the BH and accretion disk (e.g., Wilms et al.\ 
2001). 
Still, it is clear that the success of these powerful jets in making 
the extended radio lobes typical of RLQs
depends critically on the resistance the fledging jets encounter in
crossing the SMBH environment.

A jet with kinetic power $L_j$ and Lorentz factor $\Gamma$ will 
carry a mass flux 
\begin{equation}
\dot{M}_j = L_j/\Gamma c^2.
\end{equation} 
If a mass flux exceeding $\dot{M}_j$ intercepts the jet with a large
covering factor, the jet will be drastically slowed down and effectively 
quenched (e.g., Hubbard \& Blackman 2006). We investigate if a plausible 
source of thermal plasma needed for mass loading the nascent jet is 
the wide-spread debris resulting from tidal disruption of stars in the 
vicinity of the SMBH. This nuclear tidal debris (NTD), although first 
invoked 
to fuel quasars (e.g., Hills 1975), has been 
normally deemed rather inadequate (e.g., Frank \& Rees 1976).  However, 
NTD accretion is often invoked to explain the X-ray or UV flares occurring 
in the nuclei of some non-active galaxies (Rees 1988; Gezari et al.\ 2006). 
As argued by Rees (1988) and Ayal et al.\ (2000), part of the tidal 
debris would accrete on to the SMBH, either in Bondi mode, or through 
a disk (e.g., Hills 1978). There are two important time scales:
the average time interval, $t_i$, between the close approaches
of the stars to the SMBH, and the viscous 
time scale of accretion through a disk or torus, $t_v$. Rescaling the 
results of Evans \& Kochanek (1989) who carried out numerical studies 
of stellar disruption for $M_{BH}=10^6 M_{\sun}$, to the $10^8 M_{\sun}$ mass range,
we estimate that for a star of energy $E$, the orbital time, roughly
$\propto \sqrt{M_{BH}^2/(-E)^3}$ is $\sim 10^3$ times longer (i.e., hours). 
The tidal interaction energy, being $\propto M_{BH}^{2/3}$, implies  more
energetic disruptions of stars around the most massive BHs.
The viscous time is higher than the orbital time by a factor, 
${\cal M}^2 /\alpha=10^7$, where the Mach number
${\cal M}=v_\phi/ c_s$. This would yield a torus or disk life time of 
about $10^3$ yrs. Since we expect $t_i \simeq 100$ yrs (see below), 
$t_i \lesssim t_v$. This result, along with independent orbital planes of successively 
captured stars, would result in a long lasting, quasi-spherical distribution 
of debris planes, and thus a Bondi type infall of the stripped gas would 
ensue.
Since the inspiral time due to gravitational radiation losses ($\propto M_{BH}^{8/3}$) 
is orders 
of magnitude longer for the most massive 
SMBHs,  the above scenario is strengthened. 
But, to
be on the conservative side, this additional input to accretion rate, and hence
perhaps to the jet power, will be ignored here.

A crucial point for this scenario is that most estimates 
place the central BH of powerful quasars (RLQs and RQQs) in the mass range 
$\sim 10^8 - 10^9 M_{\sun}$ (\S 1). At these masses, even a factor of two 
decrease in $M_{BH}$ could have drastic consequences for NTD creation,  
rendering a jet's emergence critically dependent on $M_{BH}$.  For 
$M_{BH} = M_{c} \approx  3 \times 10^8 M_{\sun}$ the Schwarzschild radius, 
$r_s$, becomes equal to the tidal radius, $r_T$, at which solar type 
stars would be shredded upon passage close to the BH (Hills 1975).  Due to 
relativistic effects, and depending on the BH spin, 
$M_c$ can be reduced to become close to $1 \times 
10^8 M_{\sun}$ (Hills 1978; Rees 1988).  Then even if the pericenter of 
a star, $r_{min}$, becomes smaller than $r_T$ ($<r_s$) it will create 
little debris via tidal disruption, as it is swallowed whole by the SMBH. 
For slightly less close flybys ($r_{min}$ = 2--3  $r_T$) only the outer 
layers of the star are disrupted. Therefore, for $M_{BH} > M_c$, most 
main sequence stars and all compact remnants would be swallowed whole by 
the central SMBH and the only contributors left to generate debris around 
the SMBH are giant stars (Hills 1978; Frank 1978). Although rare within the 
cores of early-type galaxies, giants could be responsible for the jet abortions 
needed to explain the association of a few RQQs with extremely massive 
black holes ($\sim 3 \times 10^8 M_{\sun}$; e.g., MM). 
Thus, in our scenario, when $M_{BH} \rightarrow 
3 \times 10^8 M_{\sun}$, the disappearance of the captured typical 
(sun-like) stars into the BH event horizon, without producing jet 
disrupting NTD, can immensely boost the prospects of the jet's escaping 
from its origin to form an RLQ. For the majority of stars the differences 
in the value of $M_c$ that arise from variations in the stars' binding 
energies, will be slight ($M_c \propto \bar{\rho}_*^{-1/2}$) so that
a systematic excess of SMBH mass for RLQs should be maintained.

                                                                                
Stellar dynamical models  require that a certain fraction of stars in the 
central region of an active galaxy will pass within $r_T$ of the central 
SMBH (e.g., Frank \& Rees 1976; Rees 1988).
A good approximation to the rate at which mass is shred in the vicinity
of the BH by tidal disruption of stars is given by (Frank 1978; Rees 1988)
\begin{equation}
\dot{M}_D = 4.6\times 10^{-2}M_8^{4/3}n_{c,5} \sigma_2^{-1}(r_{min}/r_T),
\end{equation}
where $M_8$ is the mass of the SMBH in units of $10^8$M$_{\sun}$, $n_{c,5}$ 
is the density of typical (main sequence) stars in the galactic core in units 
of $10^5$pc$^{-3}$, $\sigma_2$ is the core's stellar velocity dispersion 
in units of 100 km s$^{-1}$, and $r_{min}$ is the minimum distance of 
approach to the SMBH. As the BH provides a sink, the stellar distribution 
cannot be isothermal, but follows a power-law cusp in stellar density 
(Peebles 1972; Bahcall \& Wolf 1976).  Taking into account the depletion of 
stars in low angular momentum (loss cone) orbits and replenishment by 
diffusion, Frank \& Rees (1976) found that the resulting mass loss rate 
retains the form of Eq.\ (2), and is close 
to estimates made by Hills (1975).  
Recent numerical studies (e.g., Magorrian \& Tremaine 1999; Wang \& Merritt 
2004) seem to indicate 10 -- 100 times lower feeding rates.
Although the classical loss cone theory which was developed for the BH
in centers of globular clusters is not valid for young galactic nuclei
(i.e., lower mass unrelaxed systems, treated by Wang \& Merritt 2004 and Magorrian 
\& Tremaine 1999; see above), it is quite relevant for systems with masses 
($\gtrsim 10^8 M_\odot$) which are of interest here; see Merritt (2006).  
However, all disruption rate predictions are beset with considerable 
uncertainties, summarized by Merritt (2006):  the classical loss cone 
theory depends critically on the  stellar density profiles, the system's
relaxation state, how the distribution function depends 
on angular momentum and the stellar contribution to the gravitational 
potential. All these  can significantly change the outcome.  In particular, 
in collisionless young galactic nuclei, feeding rates can be much
higher than in a relaxed nucleus if the nucleus is triaxial and many of
the orbits are ``centrophilic'' (Merritt 2006).

Direct observations, on the other hand, can only
provide a lower limit to $\dot{M}_D$, and the feeding rates 
can be an order-of-magnitude higher than the 
$10^{-3} $M$_\sun$ yr$^{-1}$ estimated by Donely et al.\ 
(2002).  For example, Ivanov et al.\ (2005) argue that the
observed flaring rate weakly constrains the disruption rate to be of order 
10 times the nominal rate. In view of these uncertainties, we adopt here
the analytical estimate given by Eq.\ (2). 
                                                                                
It is expected that about half of the tidal debris will be expelled in a fast 
moving spray of gas at a characteristic velocity $\sim 10^4 M_6^{1/6}$ 
km s$^{-1}$ (Rees 1988).  We propose that this expelled debris forms
the broad absorption line clouds (\S 3), whereas the debris bound to the BH 
would 
eventually fall back to $r \sim r_T$ and then it could be partly ejected 
in a sustained wind driven by the quasar radiation.  A substantial fraction,
$\eta$, of $\dot{M}_D$ will, however, be NTD spread in streams or clumps 
 throughout a parsec-scale region around the SMBH,
albeit concentrated in the inner few $r_T$.

The jet should have a fairly large solid angle, $\Omega_j \sim 0.3-1.0$ sr 
on sub-parsec scale, as revealed by VLBI for M87 (Biretta, Junor \& Livio 2002)
 and Cen A (Horiuchi et al.\ 2006). The tidally stripped gas 
may eventually fall in nearly isotropically or it may be funneled through
a bi-cone of solid angle $\Omega_D$ somewhat less that $2\pi$ sr aligned
with the SMBH spin axis. 
Thus, the condition for jet disruption due to mass loading by the NTD becomes
\begin{equation}
\eta \dot{M}_D \Omega_j / \dot{M}_j \Omega_D > 1.
\end{equation}
Substituting from Eqns (1) and (2) for $\dot{M}_j$ and $\dot{M}_D$, 
\begin{equation}
M_8^{4/3}n_{c,5} \sigma_{c,2}^{-1} \eta f_{0.1} L_{44}^{-1} \Gamma_{10} > 0.038,
\end{equation}
where we have conservatively taken $r_{min} = r_T$, and normalized to 
typical values $f_{0.1} = 10 \Omega_j/\Omega_D$, $\Gamma_{10} = \Gamma/10$ 
and $L_{44} = L_j/10^{44}$erg s$^{-1}$, corresponding to a moderately 
powerful jet.  In Fig.\ 1 we show the regimes in which a jet of kinetic
power $L_j$
is likely to be aborted by the NTD via mass loading, for various combinations
of $n_c$ ($10^4 - 10^8$ pc$^{-3}$) and $\Gamma$ (1--100).  Estimated values 
for $n_c \sim 10^7$ pc$^{-3}$ are indicated in several observations and 
the stellar cores have a typical radius of order of 10 pc (based on HST 
imaging by Capetti \& Balmaverde 2005); however we normalize conservatively 
for a fiducial value of $10^5$ pc$^{-3}$. Further, we take $\sigma_{c}$ = 
300 km s$^{-1}$, $\eta = 0.5$ and $f = 0.1$ as typical values,  as they
vary less than the other parameters.
Fig.\ 1 also shows two nominal upper limits to jet's kinetic
power based on the SMBH Eddington luminosity: 0.1 $L_{Edd}$, which 
is probably realistic; and a strong upper limit of 1.0 $L_{Edd}$.  

\begin{figure}
\includegraphics[angle=0,scale=0.84]{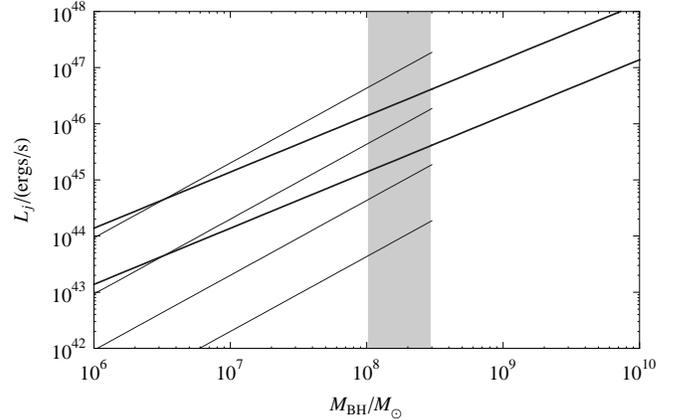} 
\caption{Allowed ranges for $L_j$ against $M_{BH}$.  The two thick
lines denote $L_{Edd}$ (top) and $0.1 L_{Edd}$ (bottom) upper limits
to the jet KINETIC luminosity.  The vertical band indicates the
range of $M_c$.  The thin lines show the minimum $L_j$ needed for
successful ejection beyond pc-scales for (top to bottom):
$n_c \Gamma = 10^8, 10^7, 10^6, 10^5$ pc$^{-3}$.
}
\end{figure}

For $M_{BH} > M_{c} = 1-3 \times 10^8$M$_\sun$, jets of any power can 
escape as the tidal debris is largely absent. But even in this regime,
if $n_c > 10^8$ pc$^{-3}$, stellar collisions will release debris 
(e.g., 
Frank 1978), but this contribution is unlikely to be relevant 
unless $M_8 \gg 1$. On the other hand, for $M_{BH} < M_c$,
Fig.\ 1 places significant lower limits to the powers of jets which can 
successfully pierce the NTD cloud. For AGN with such central SMBH and 
relatively
large $n_c$ only jets of sufficiently low $\Gamma$ can escape disruption 
by NTD. 
At the same time, since the maximum jet power attainable 
is expected to be some fraction ($\sim 0.1$) of $L_{Edd}$, 
the range of kinetic power of successful jets is constricted.


Thus, in our model, there are two regimes that allow escape of low power 
jets from the nuclear region: 1) a relatively low mass central BH,
as found in Seyfert galaxies ($M_8 \sim 0.03 - 0.3$), launching rather
slow jets through stellar cores of relatively low density; 2) 
an extremely massive BH  ($M > M_{c}$), as in RLQs (\S 1). 
The ability of even weak jets to be successfully launched by such massive 
BHs accords with the finding that RLQs of very similar $M_{BH}$ can span 
an enormous range in radio luminosity (e.g.\ MM).
This also explains why even weak AGN, such as LINERS, are able to eject 
(low power) jets if they are hosted by massive ellipticals with $M_{BH} 
> M_{c}$ (Chiaberge et al.\ 2005; Maoz 2007).

The recent study by White et al.\ (2007) applied an image stacking technique
to the FIRST survey to obtain a sample of over 41,000 SDSS quasars.
They find that the radio luminosity rises roughly as the 
0.85 power of the optical luminosity, implying that the radio loudness
factor $R$ slowly declines with optical power.  This result concurs
with our picture, in that the optically most luminous objects, expected to
host the most massive BHs, can have a wide range of (unbeamed) 
radio powers, whereas the lower mass objects only have jet powers in a
narrow range with the
 minimum set by tidal debris and the maximum imposed 
by the Eddington limit (Fig.\ 1). 

 
\section{Additional Observational Implications}

We now briefly discuss some other observational results that fit nicely within
our picture.


Although broad absorption line (BAL) gas has now been detected in  
$ > 25\%$ of quasars (e.g., Trump et al.\ 2006), 
its dearth among the powerful, edge-brightened (FR II) 
radio sources seems striking (Gregg, Becker \& de Vries 2006).
However, this anti-correlation is absent at 10--100 $\mu$Jy 
flux levels where  BAL quasars are, in fact, found to be radio brighter 
than their non-BAL counterparts (White et al.\ 2007). Attributing the 
difference to Doppler boosting, White et al.\ have suggested that the 
the jets in BALQSOs are even better aligned to our direction than the jets of 
the majority of the QSOs, which would imply that the BAL clouds 
are also present in the polar direction. They interpret this in
terms of an evolutionary unification model (L{\'i}pari \& Terlevich 2006) 
where during a specific phase, low-level radio emission still confined near
the central engine  is accompanied by an abundance of absorbing clouds that
are quickly eliminated as the radio jet breaks out to form a 
genuine RLQ. 

Alternatively, in our picture,  the jet's 
breaking out, and the consequent formation of a powerful radio source, 
is usually only possible if there is a dearth of NTD. Should the debris 
be created 
via stellar disruption (for $M_{BH} < M_c$), not only would it cause jet 
abortion though mass loading but also an appreciable fraction of the 
gaseous debris would be expelled at $\gtrsim ~ 10^4$ \kms (Rees 1988).  
In our model these gaseous clumps are observed as the BAL clouds, 
ionized in the intense radiation field of the accretion disk and its 
corona.  This is also consistent with the observed high abundance of
Fe II in BALQSOs (e.g., Yuan \& Wills 2003). 


Further, active galaxies with luminous extended emission line regions 
(EELRs), on a 10 -- 100 kpc scale, have recently been found to have lower 
(sub-solar) metallicities in their broad line regions (BLRs), as 
compared to the super-solar metallicities typically found in the BLRs 
of weak EELR galaxies (Fu \& Stockton 2007a).  
Since EELRs are thought to be blown out by jets
(e.g.\ de Breuck et al.\ 2000; Fu \& Stockton 2007b) their presence 
requires, in our scenario, a dearth of tidal debris. The lack of high 
metallicity BLR clouds can thus indeed be expected to correlate with 
jet-induced EELR prominence.

Lastly, while mergers of two SMBHs are expected to yield core type 
ellipticals (cf.\ Capetti \& Balmaverde 2006), they are unlikely to cause the 
observed high incidence of radio-loudness in such ellipticals via 
spinning up the SMBHs. Following a merger of identical SMBHs with randomly 
oriented spins, a simple vector addition averaged over all directions of
aligned angular momenta and magnetic field, $B$, results in a reduction of the spin energy 
loss rate by a factor of 12 since it is $\propto a^2 B^2 M_{BH}^2$ (here 
$B$ is normalized for flux conservation and $a$ for mass conservation). 
Taking radiation of angular momenta and relativistic effects into account, 
this rate can drop even further (Hughes \& Blandford 2003). A plausible
alternative follows from the result that SMBHs in core type ellipticals 
are typically a few times more massive than those found in power-law 
ellipticals (Capetti \& Balmaverde 2006). It is this factor (leading to 
more cases of $M_{BH}>M_c$) that is likely to explain the strong link of 
radio-loudness to core type ellipticals. The crucial role of $M_{BH}$ is 
evident from Fig. 7 of Capetti \& Balmaverde (2006) where all 
quasars with $M_{BH} > 3 \times 10^8 M_\odot$ are found to be radio loud 
(see, also, Laor 2000).


To summarize, compared to their RQQ counterparts, 
the observed systematic excess of the masses of the central BHs in RLQs
could account for the radio dichotomy, despite the mass excess being  
just by about a factor of two. This is because even this marginally larger mass 
can ensure that the tidal disruption radii for most (i.e., main sequence) stars 
fall within the event horizons of the SMBHs for most RLQs. Therefore, only in 
these more massive central engines will the tidal debris material not be 
available for mass loading the nascent jets, allowing them to emerge.
In contrast, the majority of jets launched from less massive BHs 
will encounter widespread tidal debris and will probably be disrupted 
on the sub-parsec scale, thus appearing as radio-quiet AGN.
Since part of the tidal debris is  expected to be expelled 
at high velocities ($\sim 10^4$ \kms), it could give rise to the metal-rich 
BAL features. This way, the marked lack of BALs in extended radio quasars 
also can be easily understood.



\acknowledgments
We thank the referee for insightful comments.        
PJWs work is supported in part by a GSU subcontract to NSF grant AST05-07529 at 
the University of Washington.

\end{document}